\documentclass[aps,prd,preprintnumbers,superscriptaddress,nofootinbib,amsmath,twocolumn,amssymb,showpacs]{revtex4}%,twocolumn showpacs
\usepackage{graphicx}% Include figure files
\usepackage{dcolumn}% Align table columns on decimal point
\usepackage{bm}% bold math
\usepackage{color}
\input{colordvi.tex}
\newcommand{\beq}{\begin{equation}}
\newcommand{\eeq}{\end{equation}}
\newcommand{\beqa}{\begin{eqnarray}}
\newcommand{\eeqa}{\end{eqnarray}}
\newcommand{\mpl}{M_{\rm Pl}}

\newcommand{\GeV}{~\mbox{\rm GeV}}
\newcommand{\cH}{{\cal H}}

\newcommand{\D}{{\rm d}}
\begin{document}

\title{Generalized Higgs inflation}

\author{Kohei Kamada}
\email[Email: ]{kohei.kamada"at"desy.de}
\affiliation{Deutsches Elektronen-Synchrotron DESY,
Notkestrasse 85, D-22607 Hamburg, Germany}

\author{Tsutomu~Kobayashi\footnote{Present address: Department of Physics, Rikkyo University, Toshima,
Tokyo 171-8501, Japan.}}

\email[Email: ]{tsutomu"at"tap.scphys.kyoto-u.ac.jp}
\affiliation{Hakubi Center, Kyoto University, Kyoto 606-8302, Japan
}
\affiliation{Department of Physics, Kyoto University, Kyoto 606-8502, Japan}

\author{Tomo Takahashi}
\email[Email: ]{tomot"at"cc.saga-u.ac.jp}
\affiliation{Department of Physics, Saga University, Saga 840-8502, Japan}

\author{Masahide~Yamaguchi}
\email[Email: ]{gucci"at"phys.titech.ac.jp}
\affiliation{Department of Physics, Tokyo Institute of Technology, Tokyo
152-8551, Japan}

\author{Jun'ichi~Yokoyama}
\email[Email: ]{yokoyama"at"resceu.s.u-tokyo.ac.jp}
\affiliation{Research Center for the Early Universe (RESCEU), Graduate
School of Science, The University of Tokyo, Tokyo 113-0033, Japan}
\affiliation{Kavli Institute for the Physics and Mathematics of the Universe
(IPMU), The University of Tokyo, Kashiwa, Chiba, 277-8568, Japan}

\preprint{DESY~12-043,~RESCEU-4/12}%,~KUNS-****}

\begin{abstract}
We study Higgs inflation in the context of generalized G-inflation, {\em
i.e.,} the most general single-field inflation model with second-order
field equations.  The four variants of Higgs inflation proposed so far
in the literature can be accommodated at one time in our framework.  We
also propose yet another class of Higgs inflation, {\em the running
Einstein inflation model}, that can naturally arise from the generalized
G-inflation framework.  As a result, five Higgs inflation models in all
should be discussed on an equal footing. Concise formulas for primordial
fluctuations in these generalized Higgs inflation models are provided,
which will be helpful to determine which model is favored from the
future experiments and observations such as the Large Hadron Collider
and the Planck satellite.
\end{abstract}
\maketitle

\section{Introduction}

The Higgs particle is the only undiscovered ingredient of the standard
model (SM) of particle physics; it plays the fundamental role of
accounting for the origin of the masses of all the known massive
particles. Though some signals have been hinted at in the 
LHC experiments recently {\cite{ATLAS,CMS}}, we awaited its final discovery. 
The discovery of the Higgs particle would have profound
implications not only in particle physics but also in cosmology, since
all of the inflation models rely on the existence of a scalar field, the
inflaton, driven either by its potential energy
\cite{Sato:1980yn,Linde:1983gd} or kinetic energy
\cite{kinflation,G-inf}.  Note that even higher-curvature theories of
inflation without any scalar field
\cite{Starobinsky:1980te,hep-ph/9807482} may be conformally transformed
to Einstein gravity with a scalar field driving inflation.

There may even be a direct connection between the Higgs field and cosmic
inflation, namely, the possibility that the Higgs field itself acts as
the inflaton. In order to suppress the amplitude of the curvature
perturbation from the inflaton's quantum fluctuations \cite{fluctuation},
its self-coupling $\lambda$ must be smaller than $\sim 10^{-13}$
\cite{DAMTP-R-92-26}, which is not the case in the SM Higgs field.
Hence, some extension is necessary in either gravitational or kinetic
sectors of the theory.

So far, four variants of Higgs inflation have been proposed in this
direction.\footnote{Inflationary models in which the Higgs field in supersymmetric 
standard models is identified as the inflaton
are discussed in Ref.~\cite{Einhorn:2009bh}. } The first one is to introduce a large and negative
nonminimal coupling between the scalar field and the scalar curvature
\cite{Spokoiny,FM,nonmin}. In this model, the Planck scale takes
effectively a much larger value during inflation than it is today to
suppress the amplitude of curvature perturbations.

The second is the new Higgs inflation model \cite{nonder} whose kinetic
term is coupled to the Einstein tensor \cite{arXiv:1104.2253}.  This
coupling changes the normalization of the field during inflation, which
suppresses quantum fluctuations.  The third one is running kinetic
inflation~\cite{Nakayama-Takahashi}, in which a nonstandard kinetic
term simply changes the normalization of the inflaton in some domain of
the field space, leading essentially to the same effect as in the
previous example.  Finally, the fourth model is Higgs
G-inflation~\cite{arXiv:1012.4238}, where the lowest nontrivial-order
Galileon-like interaction~\cite{G1,G2} is incorporated into the original
action.  Although this model contains higher-derivative interactions,
the field equations remain of second order and the newly introduced term
acts as an extra friction, which effectively smoothens the potential to
suppress curvature fluctuations down to the observed value.

In fact, each of the above four models falls into a subclass of
generalized G-inflation~\cite{arXiv:1105.5723}, which is the most
general single-field inflation model having second-order gravitational
and scalar-field equations. Hence, a unified treatment of apparently
different Higgs inflation models is possible in the context of
generalized G-inflation. As a by-product of this fact, we propose yet
another class of successful Higgs inflation.

In this paper, we first clarify why five different Higgs inflation
models exist in the context of generalized G-inflation. Then, we discuss
their dynamics and primordial fluctuations in a unified way. In
particular, the formulas of primordial fluctuations in these generalized
Higgs inflation models are given in terms of the slow-roll parameters
and field-dependent functions in the Lagrangian, which will be helpful
to single out the model favored by the future experimental and
observational data from the LHC experiment and the Planck satellite, 
etc. Note, however, that in the context of generalized G-inflation, one
may well find the best-fit model in some combinations of two or more
models among the five mentioned above.  Indeed, the strength of the
generalized G-inflation is that, in performing the parameter search using
the Markov chain Monte Carlo (MCMC) method, all the variants of Higgs
inflation models can be analyzed simultaneously and seamlessly unlike in
\cite{arXiv:1107.3436}.

The purpose of the present paper is therefore to clarify first how the
previously known four Higgs inflation models are realized as part of
the generalized G-inflation model and then propose the fifth model in
the same context together with the formulas for curvature perturbations and 
tensor perturbations, as well as the non-Gaussianity of the former,
which turns out to be small, in a unified manner.  Note that our
framework is not confined only to inflation driven by the SM Higgs field
but is applicable to more general potential-driven single-field
inflation models, too.

The rest of the paper is organized as follows. In Sec. II, we introduce
variants of Higgs inflation models in the context of generalized
G-inflation. In Sec. III evolution of the homogeneous background and
conditions for inflation are summarized.  Then we calculate spectra of
perturbations in Sec. IV in a unified manner.  Finally, Sec. V is devoted to
a discussion and conclusions.

\section{Higgs inflation models as variants of generalized $G$-inflation}

The tree-level SM Higgs Lagrangian is
\begin{equation}
S_0=\int \D^4 x \sqrt{-g} \left[\frac{\mpl^2}{2}R - |D_\mu {\cal H}|^2-\lambda(|{\cal H}|^2-v^2)^2 \right], 
\end{equation}
where $\mpl$ is the reduced Planck mass, $D_\mu$ is the covariant
derivative with respect to the SM gauge symmetry, ${\cal H}$ is the SM
Higgs boson, $v$ is its vacuum expectation value, and $\lambda$ is
the self-coupling constant.  Taking the gauge
$^t\cH=(0,v+\phi)/\sqrt{2}$, {with $\phi$ being a real scalar field} and
assuming $\phi \gg v$, the action is simplified to
\begin{equation}
S_0=\int \D^4 x \sqrt{-g} \left[\frac{\mpl^2}{2}R - \frac{1}{2}(\partial_\mu \phi)^2-\frac{\lambda}{4}\phi^4\right], 
\end{equation}
which is nothing but the action for original chaotic inflation~\cite{Linde:1983gd}.

This model cannot serve as a viable inflation model as it stands.
Since the self-coupling is related with the Higgs mass $m_H$ as
\begin{equation}
  m_H=\sqrt{2\lambda}v,\quad v=246\GeV,
\end{equation}
at the tree level,  $\lambda$ cannot take a tiny value to give the correct amplitude for density fluctuations
with the value indicated by the LEP collider  $m_H>114.4$ GeV at the 95\% CL \cite{Barate:2003sz}.

As mentioned in the Introduction, four remedies have been proposed so
far, all of which can be unified 
as a subclass of the generalized G-inflation~\cite{arXiv:1105.5723} whose
action is given by
\begin{eqnarray}
S=\sum_{i=2}^{5}\int{\rm d}^4x\sqrt{-g}{\cal L}_i,
\end{eqnarray}
where
\begin{eqnarray}
{\cal L}_2&=&K(\phi, X),\label{L2}
\\
{\cal L}_3&=&-G_3(\phi, X)\Box\phi,
\\
{\cal L}_4&=&G_{4}(\phi, X)R+G_{4X}\left[
\left(\Box\phi\right)^2-\left(\nabla_\mu\nabla_\nu\phi\right)^2
\right],
\\
{\cal L}_5&=&G_5(\phi, X) G_{\mu\nu}\nabla^\mu\nabla^\nu\phi
-\frac{1}{6}G_{5X}\Bigl[
\left(\Box\phi\right)^3
\nonumber\\&&
-3\left(\Box\phi\right)\left(\nabla_\mu\nabla_\nu\phi\right)^2
+2\left(\nabla_\mu\nabla_\nu\phi\right)^3
\Bigr],\label{L5}
\end{eqnarray}
where $R$ is the Ricci tensor, $G_{\mu\nu}$ is the Einstein tensor,
$X=-(1/2)g^{\mu\nu}\nabla_\mu\phi\nabla_\nu\phi$,
$(\nabla_\mu\nabla_\nu\phi)^2=\nabla_\mu\nabla_\nu\phi\nabla^\mu\nabla^\nu\phi$,
$(\nabla_\mu\nabla_\nu\phi)^3=\nabla_\mu\nabla_\nu\phi\nabla^\nu\nabla^\lambda\phi\nabla_\lambda\nabla^\mu\phi$,
and $G_{iX}=\partial G_i/\partial X$.
This theory was originally discovered by Horndeski \cite{Horndeski}
in a different form, and rediscovered by Deffayet et al.\ \cite{Deffayet:2011gz}
in the present form, whose equivalence to the original theory was
first shown in \cite{arXiv:1105.5723}.

For a homogeneous and isotropic cosmological background,
$\D s^2=-\D t^2+a^2(t)\D\mathbf{x}^2$, $\phi=\phi(t)$,
the $(tt)$ component of the gravitational field equations reads
\begin{eqnarray}
\sum_{i=2}^5{\cal E}_i = 0, \label{E}
\end{eqnarray}
where
\begin{eqnarray}
{\cal E}_2&=&2XK_X-K,
\\
{\cal E}_3&=&6X\dot\phi HG_{3X}-2XG_{3\phi},
\\
{\cal E}_4&=&-6H^2G_4+24H^2X(G_{4X}+XG_{4XX})
\nonumber\\&&
-12HX\dot\phi G_{4\phi X}-6H\dot\phi G_{4\phi },
\\
{\cal E}_5&=&2H^3X\dot\phi\left(5G_{5X}+2XG_{5XX}\right)
\nonumber\\&&
-6H^2X\left(3G_{5\phi}+2XG_{5\phi X}\right),
\end{eqnarray}
with $H=\dot a/a=\D \ln a/\D t$.
This corresponds to the Friedmann equation, which can be easily verified 
by substituting $G_4=\mpl^2/2=$const, and 
$G_3=0=G_5$
into the above equations.
The scalar-field equation of motion is given by
\begin{eqnarray}
\frac{1}{a^3}\frac{\rm d}{{\rm d} t}\left(a^3 J\right) =P_\phi,\label{phieom}
\end{eqnarray}
where
\begin{eqnarray}
J&=&\dot\phi K_X+6HXG_{3X}-2\dot\phi G_{3\phi}
\nonumber\\&&
+6H^2\dot\phi\left(G_{4X}+2XG_{4XX}\right)-12HXG_{4\phi X}
\nonumber\\&&
+2H^3X\left(3G_{5X}+2XG_{5XX}\right)
\nonumber\\&&
-6H^2\dot\phi\left(G_{5\phi}+XG_{5\phi X}\right), 
\end{eqnarray}
and
\begin{eqnarray}
P_\phi &=&
K_\phi-2X\left(G_{3\phi\phi}+\ddot\phi G_{3\phi X}\right)
\nonumber\\&&
+6\left(2H^2+\dot H\right)G_{4\phi}
+6H\left(\dot X+2HX\right)G_{4\phi X}
\nonumber\\&&
-6H^2XG_{5\phi\phi}+2H^3X\dot\phi G_{5\phi X}.
\end{eqnarray}
The space-space component of the gravitational field equations
is not independent of the generalized
Friedmann and scalar-field equations.

Although the  generalized G-inflation covers
all the possible single-field inflation models
including the ones driven by $\phi$'s kinetic energy,
since we are interested in potential-driven inflation here,
we focus on its subclass 
for which each function in the Lagrangian can be Taylor-expanded in
terms of $X$ as
\begin{eqnarray}
K(\phi, X)&=&-V(\phi)+{\cal K}(\phi)X+\cdots,
\\
G_i(\phi, X)&=&g_i(\phi)+h_i(\phi)X+\cdots.
\end{eqnarray}
Hereafter, we will neglect all the higher order terms in $X$.
Using this Taylor-expanded form, one can handle a vast class of
potential-driven inflation models while avoiding the situation
where the equations are too general to tell anything concrete.

We note here the following identities:
\begin{eqnarray}
g_3(\phi)\Box\phi& =& 2g_3'X+({\rm t.d.}),
\\
g_5(\phi)G^{\mu\nu}\nabla_\mu\nabla_\nu\phi
&=&
-g_5'\left[XR+(\Box\phi)^2-(\nabla_\mu\nabla_\nu\phi)^2\right]
\nonumber\\&&+3g_5''X\Box\phi-2g_5'''X^2+({\rm t.d.}),
\end{eqnarray}
where
a prime denotes differentiation with respect to $\phi$ and
(t.d.) represents  total derivative terms.
These identities allow us
to set $g_3=0=g_5$ without loss of generality.
In particular, the derivative coupling to the Einstein tensor
in new Higgs inflation, $G^{\mu\nu}\partial_\mu\phi\partial_\nu\phi$,
is obtained most straightforwardly from ${\cal L}_5=-\phi G^{\mu\nu}\nabla_\mu\nabla_\nu\phi$,
but that interaction
can also be obtained equivalently from
${\cal L}_4=XR+(\Box\phi)^2-(\nabla_\mu\nabla_\nu\phi)^2$.
We choose to employ the latter expression for new Higgs inflation.
Hereafter, we write $g_4 =g$.

The four remedies of Higgs inflation
proposed so far can be reproduced by
adding the extra term $\Delta {\cal L}$ to the standard Lagrangian,
$\mpl^2 R/2+X-V(\phi)$, where $\Delta {\cal L}$ is given respectively by
\begin{eqnarray}
\Delta {\cal L}&=&\kappa\phi^{ {2}n} X
\quad
{\rm (running~kinetic~inflation)},
\\
\Delta {\cal L}&=&\frac{\phi }{M^4}X\Box\phi
\quad
{\rm (Higgs~G\mathchar`-inflation)},
\\
\Delta{\cal L}&=&-\frac{\xi}{2}\phi^2 R
\quad
{\rm (non\mathchar`-minimal~Higgs~inflation)},
\end{eqnarray}
and
\begin{eqnarray}
\Delta{\cal L}&=&
\frac{1}{2\mu^2}\left[X R+(\Box\phi)^2-(\nabla_\mu\nabla_\nu\phi)^2\right]
\nonumber\\
&&\qquad{\rm (new~ Higgs~ inflation)}.
\end{eqnarray}
Here  $\kappa$ and $\xi$ are dimensionless constants,
and $M$ and $\mu$ are parameters having dimension of mass.
All of those apparently different models can be treated in a unified 
manner by taking
\begin{eqnarray}
  {\cal K}(\phi) &=& 1+\kappa\phi^{2n}, \\
  h_3(\phi) &=&  \frac{\phi}{M^4},\\
  g(\phi) &=& \frac{\mpl^2}{2}-\frac{\xi}{2}\phi^2,\\
  h_4(\phi) &=&  \frac{1}{2\mu^2},
\\   h_5(\phi)&=&0.
\end{eqnarray}

It is then natural to imagine the case with $h_5(\phi)\neq
0$,\footnote{The simplest example of $h_5(\phi)$ for Higgs
inflation is $h_5(\phi) = \phi / \Lambda^6$ with $\Lambda$ being some
cutoff scale. Note also that, in order to guarantee the gauge invariance 
of the Higgs doublet, the power of $\phi$ in ${\cal K}, g$ and $h_4$ must be even, 
while that in $h_3$ and $h_5$ must be odd.}  which would lead to yet another
successful Higgs inflation model that has not been explored before. We
call it {\it running Einstein inflation}, since it is supported by the
change of the coefficient of the Einstein tensor.

In the following analysis, we will consider those all five
possibilities of Higgs inflation on equal footing,
by characterizing  potential-driven inflation in terms of
the five arbitrary functions of $\phi$, $ {\cal
K}, g, h_3, h_4, h_5$, besides the potential $V$.

\section{General slow-roll dynamics of potential-driven inflation}

In order to investigate the general slow-roll dynamics of
potential-driven inflation including the variants of Higgs inflation, we
assume the following slow-roll conditions,
\begin{eqnarray}
&&\epsilon:=-\frac{\dot H}{H^2}\ll 1,
\quad
\eta:=-\frac{\ddot\phi}{H\dot\phi}\ll 1,
\quad
\delta:=\frac{\dot g}{Hg}\ll 1,
\nonumber\\&&
\alpha_2:=\frac{\dot{\cal K}}{H{\cal K}}\ll 1,\quad
\alpha_i:=\frac{\dot h_i}{Hh_i}\ll 1\;\;(i=3,4,5).
\end{eqnarray}
We also assume that $\dot\delta/H\delta, \dot\alpha_i/H\alpha_i\ll 1\;(i=2,3,4,5)$.
It is then found that
\begin{eqnarray}
J\simeq {\cal K}\dot\phi+3h_3H\dot\phi^2+6h_4H^2\dot\phi+3h_5H^3\dot\phi^2,
\label{s-r-J}
\end{eqnarray}
and the slow-roll equation of motion for the inflaton is given by
\begin{eqnarray}
3HJ \simeq -V'+12H^2g'.\label{sreom1}
\end{eqnarray}
This is the generalized slow-roll equation of motion for $\phi$, where
we can see how each term in Eq.~(\ref{s-r-J}) modifies the structure of
the friction term.  We can also see that the nonminimal coupling in $g$
changes effectively the slope of the potential.

We are considering potential-dominated inflation, so that $V\gg{\cal
O}(\dot \phi J)$.  Then, the gravitational field equations read
\begin{eqnarray}
6gH^2&\simeq&V,\label{srfr}
\\
-4g\dot H+2g'\dot\phi H&\simeq&\dot\phi J.
\end{eqnarray}
The second equation can be derived from Eqs.~(\ref{sreom1})
and~(\ref{srfr}), or more directly from the space-space component of the
gravitational field equations.  From the Friedmann equation~(\ref{srfr})
one can see that $2g$ may be regarded as an effective Planck mass
squared.  We should only consider the domain $g>0$ \cite{FM}, which is
always satisfied in the nonminimal Higgs inflation model since it
adopts a large and negative $\xi$.  Using Eq.~(\ref{srfr}), one can
remove $H$ from the right hand side of Eq.~(\ref{sreom1}) to give
\begin{eqnarray}
3HJ\simeq -g^2\left(\frac{V}{g^2}\right)'=:-U'(\phi).\label{sreom2}
\end{eqnarray}
The effective potential $U$ coincides with that introduced in
Refs. \cite{Chiba:2008ia} to derive the slow-roll conditions in the
Jordan frame.

Let us define
\begin{eqnarray}
u(\phi):={\cal K}+\frac{h_4V}{g},
\quad
v(\phi):=h_3+\frac{h_5V}{6g}.
\end{eqnarray}
Note that $u$ and $v$ are functions of $\phi$ only
and are determined completely through the functions in the Lagrangian.
Among the six functions of $\phi$
in the Lagrangian, the above particular combinations
$u(\phi)$ and $v(\phi)$
are crucial for the slow-roll dynamics and the spectra of primordial fluctuations.
In terms of $u$ and $v$, $J$ can be written as
\begin{eqnarray}
J=u\dot\phi+6HX v. \label{Juv}
\end{eqnarray}
Plugging this expression into Eq.~(\ref{sreom2})
and solving for $\dot\phi$, we get
\begin{eqnarray}
3H\dot\phi&\simeq&\frac{1}{2v}\left(-u+\sqrt{u^2-4U'v}\right).\label{sreom3}
\end{eqnarray}
Comparing this with the original equation~(\ref{sreom2}),
we find
\begin{eqnarray}
\frac{J}{\dot\phi}&\simeq&\frac{1}{2}\left(u+\sqrt{u^2-4U'v}\right)=:W(\phi).
\label{Joverdp}
\end{eqnarray}
We require $u^2-4U'v>0$ so that
Eqs.~(\ref{sreom3}) and~(\ref{Joverdp}) make sense.
In addition, it may be reasonable to assume that $u>0$.
We then have $0<u/W<2$. The consequences of this inequality
will be discussed further
in relation to the stability against linear perturbations
in the next section.

Combining Eqs.~(\ref{srfr}) and~(\ref{sreom3}), we arrive at
\begin{eqnarray}
\frac{\D\phi}{\D {\cal N}} = \frac{\dot{\phi}}{H} \simeq -2\frac{gU'}{VW},
\label{efolds}
\end{eqnarray}
where ${\cal N}:=\ln a$ is the number of e-folds.
Note that the right hand side is expressed solely in terms of $\phi$
and reduces to $-\mpl^2 V'/V$ in the case of the standard canonical field.
In general slow-roll inflation,
the effective potential slope
$2gU'/VW$ governs the motion of $\phi$
rather than the ``bare'' one $\mpl^2V'/V$.
For instance, slow roll of $\phi$ is possible even in a steep potential
if $W\gg 1$.
Equation~(\ref{efolds}) can be used to evaluate
the number of e-folds until the end of inflation.

Using Eq.~(\ref{efolds}),
each slow-roll parameter can be expressed in terms of the
$\phi$-dependent functions as
\begin{eqnarray}
\delta&\simeq&-2\frac{g'U'}{WV},
\\
\epsilon&\simeq&\frac{g}{W}\left(\frac{U'}{V}\right)^2-\frac{\delta}{2},\label{ep-del}
\\
\frac{\dot J}{HJ}&\simeq& -2\frac{g}{W}\frac{U''}{V}+\epsilon,\label{dJHJ}
\\
\alpha_i&\simeq&-2\frac{gU'}{VW}\frac{h_i'}{h_i}.
\end{eqnarray}
Note that Eq.~(\ref{ep-del}), together with $g>0$ and $W>0$,
ensures 
\begin{equation}
\epsilon+\delta/2>0. \label{epdel}
\end{equation}

The ratio $\dot\phi J/V$ can be expressed in terms of the slow-roll parameters as
\begin{eqnarray}
\frac{\dot\phi J}{V}\simeq \frac{2}{3}\epsilon+\frac{1}{3}\delta\ll 1.
\label{supp}
\end{eqnarray}
From this, the initial assumption that the potential
is dominant in the Friedmann equation is found to be consistent.

It is instructive here to demonstrate the extreme case
where only $h_5$ is nontrivial corresponding to
the running Einstein inflation model we are proposing in this paper.
In this case $U'=V'$, $u=1$,
and $v= h_5(\phi) V(\phi)/3\mpl^2$.
Noting that $\epsilon = \epsilon_{\rm std}/W$,
where $\epsilon_{\rm std}:=(\mpl^2/2)(V'/V)^2$
is the standard slow-roll parameter defined in terms of the potential,
we see that inflation proceeds even with a steep potential provided $W\gg 1$.
This occurs in the domain where $|h_5 V'V/\mpl^2|\gg 1$ is satisfied.

\section{Cosmological Perturbations in generalized Higgs inflation}

In this section, we study cosmological perturbations in generalized
Higgs inflation and present useful formulas for the spectra of tensor
and scalar perturbations.  A generic formulation of cosmological
perturbations in the most general single-field inflation model was
already given in Ref.~\cite{arXiv:1105.5723}.  For completeness, we
begin with duplicating the general formulas, and then illustrate how
they can be applied to the potential-driven models.

\subsection{Generic formulation for linear perturbations}

It is convenient to write the perturbed metric in the Arnowitt-Deser-Misner form as
\begin{eqnarray}
\D s^2=-N^2\D t^2+\gamma_{ij}\left(\D x^i +N^i\D t\right)\left(\D x^j +N^j\D t\right),
\end{eqnarray}
where
\begin{eqnarray}
&&N=1+\delta n,\quad
N_i=\partial_i\chi,\nonumber\\
&&\gamma_{ij} =a^2(t)e^{2\zeta}\left(\delta_{ij}+h_{ij}+\frac{1}{2}h_{ik}h_{kj}\right).
\end{eqnarray}
Here, $\delta n$, $\chi$, and $\zeta$ are scalar perturbations and
$h_{ij}$ is a tensor perturbation satisfying $h_{ii}=0=h_{ij,j}$.  We
choose the unitary gauge in which $\phi(t,\mathbf{x})=\phi(t)$.
Substituting the metric to the action and expanding it to second order
in perturbations, we obtain the quadratic actions for the tensor and
scalar perturbations.  For the scalar perturbations, one may use the
constraint equations to remove $\delta n$ and $\chi$ to get the
quadratic action in terms of the single variable $\zeta$.

The quadratic action for the tensor perturbations is given by
\begin{eqnarray}
S_T^{(2)} =\frac{1}{8}\int\D t\D^3x\,a^3\left[
{\cal G}_T\dot h_{ij}^2-\frac{{\cal F}_T}{a^2}
(\Vec{\nabla} h_{ij})^2\right], \label{tensoraction}
\end{eqnarray}
where
\begin{eqnarray}
{\cal F}_T&:=&2\left[G_4
-X\left( \ddot\phi G_{5X}+G_{5\phi}\right)\right],
\\
{\cal G}_T&:=&2\left[G_4-2 XG_{4X}
-X\left(H\dot\phi G_{5X} -G_{5\phi}\right)\right].
\end{eqnarray}
The squared sound speed is given by
$
c_T^2= {\cal F}_T/{\cal G}_T.
$
It is manifest from the action~(\ref{tensoraction}) that ghost and
gradient instabilities are avoided provided that
\begin{eqnarray}
{\cal F}_T>0,\quad {\cal G}_T>0.
\end{eqnarray}
Following the standard quantization procedure, the power spectrum of
the primordial tensor perturbations is found to be
\begin{eqnarray}
{\cal P}_T=8\gamma_T\left.
\frac{{\cal G}_{T}^{1/2}}{{\cal F}_{T}^{3/2}}
\frac{H^2}{4\pi^2}\right|_{\rm sound~horizon~exit},
\end{eqnarray}
where $\gamma_T=2^{2\nu_T -3}|\Gamma(\nu_T)/\Gamma(3/2)|^2(1-\epsilon
-f_T/2+g_T/2)$.
We emphasize that the power spectrum is evaluated at
sound horizon exit, because the propagation speed of
the tensor mode does not coincide with that of light in general.
Here, we have assumed that $\epsilon:=-\dot H/H^2\simeq\;$const,
\begin{eqnarray}
f_{T}:=\frac{\dot{{\cal F}}_T}{H{\cal F}_T}\simeq {\rm const},
~~{\rm and}~~
g_{T}:=\frac{\dot{{\cal G}}_T}{H{\cal G}_T}\simeq {\rm const},
\end{eqnarray}
and defined
\begin{eqnarray}
\nu_T:=\frac{3-\epsilon+g_T}{2-2\epsilon-f_T+g_T}.
\end{eqnarray}
The tensor spectral tilt is evaluated as
\begin{eqnarray}
n_T=3-2\nu_T .
\end{eqnarray}

On the other hand, the quadratic action for the scalar perturbations is
given by
\begin{eqnarray}
S^{(2)}_S=\int\D t\D^3 x\,a^3\left[
{\cal G}_S
\dot\zeta^2
-\frac{{\cal F}_S}{a^2}
(\Vec{\nabla}\zeta)^2
\right]\label{scalar2},
\end{eqnarray}
where
\begin{eqnarray}
{\cal F}_S&:=&\frac{1}{a}\frac{\D}{\D t}\left(\frac{a}{\Theta}{\cal G}_T^2\right)
-{\cal F}_T,
\\
{\cal G}_S&:=&\frac{\Sigma }{\Theta^2}{\cal G}_T^2+3{\cal G}_T,
\end{eqnarray}
and $\Sigma$ and $\Theta$ are defined as
\begin{eqnarray}
\Sigma&:=&XK_X+2X^2K_{XX}+12H\dot\phi XG_{3X}
\nonumber\\&&
+6H\dot\phi X^2G_{3XX}
-2XG_{3\phi}-2X^2G_{3\phi X}-6H^2G_4
\nonumber\\&&
+6\Bigl[H^2\left(7XG_{4X}+16X^2G_{4XX}+4X^3G_{4XXX}\right)
\nonumber\\&&
-H\dot\phi\left(G_{4\phi}+5XG_{4\phi X}+2X^2G_{4\phi XX}\right)
\Bigr]
\nonumber\\&&
+30H^3\dot\phi XG_{5X}+26H^3\dot\phi X^2G_{5XX}
\nonumber\\&&
+4H^3\dot\phi X^3G_{5XXX}
-6H^2X\bigl(6G_{5\phi}
\nonumber\\&&
+9XG_{5\phi X}+2 X^2G_{5\phi XX}\bigr),
\\
\Theta&:=&-\dot\phi XG_{3X}+
2HG_4-8HXG_{4X}
\nonumber\\&&
-8HX^2G_{4XX}+\dot\phi G_{4\phi}+2X\dot\phi G_{4\phi X}
\nonumber\\&&
-H^2\dot\phi\left(5XG_{5X}+2X^2G_{5XX}\right)
\nonumber\\&&
+2HX\left(3G_{5\phi}+2XG_{5\phi X}\right).
\end{eqnarray}

The squared sound speed of the curvature perturbations is given by
$c_S^2={\cal F}_S/{\cal G}_S$, and ghost and gradient instabilities are
avoided provided that the following conditions are satisfied:
\begin{eqnarray}
{\cal F}_S>0\quad {\cal G}_S>0.
\end{eqnarray}
As is the case of the tensor perturbations, the power spectrum of the
scalar perturbations can be easily computed as
\begin{eqnarray}
{\cal P}_\zeta =\frac{\gamma_S}{2}\left.
\frac{{\cal G}_S^{1/2}}{{\cal F}_S^{3/2}}\frac{H^2}{4\pi^2}\right|_{\rm
sound~horizon~exit},
\end{eqnarray}
where $\gamma_S=2^{2\nu_S -3}|\Gamma(\nu_S)/\Gamma(3/2)|^2(1-\epsilon
-f_S/2+g_S/2)$.
Note that the (sound) horizon crossing time for $\zeta$
is different from that for $h_{ij}$ in general.
We have assumed that $\epsilon\simeq\;$const,
\begin{eqnarray}
f_S:=\frac{\dot{\cal F}_S}{H{\cal F}_S}\simeq{\rm const},\quad
g_S:=\frac{\dot{\cal G}_S}{H{\cal G}_S}\simeq{\rm const},
\end{eqnarray}
and also define
\begin{eqnarray}
\nu_S:=\frac{3-\epsilon+g_S}{2-2\epsilon-f_S+g_S}.
\end{eqnarray}
The scalar spectral index is computed as
\begin{eqnarray}
n_s-1=3-2\nu_S.
\end{eqnarray}

\subsection{Primordial perturbations in generalized Higgs inflation}

Now we are in a position to derive concise and useful formulas for tensor
and scalar fluctuations in generalized Higgs inflation.  The four
important functions in the quadratic actions are evaluated as
\begin{eqnarray}
{\cal F}_T\simeq{\cal G}_T\simeq 2g,
\end{eqnarray}
and
\begin{eqnarray}
{\cal F}_S&\simeq&\frac{X}{H^2}u+\frac{4\dot\phi X}{H}v,
\\
{\cal G}_S&\simeq&\frac{X}{H^2}u+\frac{6\dot\phi X}{H}v.
\end{eqnarray}
It is convenient to rewrite ${\cal F}_S$ and ${\cal G}_S$ as
\begin{eqnarray}
{\cal F}_S&\simeq&\frac{g}{3}(2\epsilon+\delta)\left(4-\frac{u}{W}\right),
\\
{\cal G}_S&\simeq& g(2\epsilon+\delta)\left(2-\frac{u}{W}\right),
\end{eqnarray}
where we used Eqs.~(\ref{srfr}),~(\ref{Juv}),~(\ref{Joverdp}),
and~(\ref{supp}).  We see that ${\cal F}_T,\;{\cal G}_T>0$ since we are
assuming that the effective Planck mass squared $g$ is positive.  It
should be noted that ${\cal F}_S>0$ and ${\cal G}_S>0$ are also
guaranteed by the inequalities $\epsilon+\delta/2>0$ and $u/W<2$ which
we discussed in the previous section. The sound speed squared is given
by
\begin{equation}
  c_s^2 = \frac{4-u/W}{3(2-u/W)}.
\end{equation}
We see that $2/3\le c_s^2<\infty$, though the superluminal
propagation leads to the absence of the Lorentz invariant UV
completion \cite{Adams:2006sv}.

The tensor power spectrum is simply given by
\begin{eqnarray}
{\cal P}_T\simeq \frac{H^2}{\pi^2 g} \simeq \frac{V}{6\pi^2 g^2},
\end{eqnarray}
and its tilt is
\begin{eqnarray}
n_T\simeq -2\epsilon-\delta.
\end{eqnarray}
From (\ref{epdel}) we find it is always negative in the potential-driven
models under consideration, although the blue tensor spectrum is possible in
kinetically driven G-inflation~\cite{G-inf}.

The power spectrum of the curvature perturbations is expressed as
\begin{eqnarray}
{\cal P}_\zeta=\frac{\sqrt{3}}{16\pi^2}\frac{V}{g^2(2\epsilon +\delta)}
\frac{(2-u/W)^{1/2}}{(4-u/W)^{3/2}},
\end{eqnarray}
and the spectral index is
\begin{eqnarray}
n_s-1&\simeq& -4\epsilon+\eta-\frac{\dot J}{HJ}
   \nonumber\\&&
   +2 \frac{gU'}{VW} \left[
   \frac12 \frac{\left( u/W \right)'}{\left( 2-u/W \right)}
  -\frac32 \frac{\left( u/W\right)'}{\left( 4-u/W \right)}
   \right].
\end{eqnarray}
Thus, the tensor-to-scalar ratio is given by
\begin{eqnarray}
r&=&-\frac{8}{3\sqrt{3}}\frac{(4-u/W)^{3/2}}{(2-u/W)^{1/2}}n_T
\nonumber\\
&=&-\frac{8}{\sqrt{3}}(4-u/W)^{1/2}c_sn_T.
\end{eqnarray}
It is interesting to note that the tensor-to-scalar ratio is enhanced if
the inflaton trajectory satisfies $u/W\approx 2$, {\em i.e.,}
$u^2\approx 4U'v$.

Let us then consider two extreme cases where
\begin{eqnarray}
J\simeq u\dot\phi \quad{\rm and}\quad J\simeq 6HXv.
\end{eqnarray}
The former corresponds to $u \simeq W$,
which is the case in running kinetic inflation and new Higgs inflation, 
and the latter to $u \ll W$, which is the case in Higgs G-inflation
and running Einstein inflation.
In both limiting cases we have ${\cal F}_S, {\cal G}_S\propto
(\dot\phi /H^2)J$, so that
\begin{eqnarray}
n_s-1\simeq -4\epsilon+\eta-\frac{\dot J}{HJ}.\label{ns-extreme}
\end{eqnarray}

If $J\simeq u\dot\phi$,
the power spectrum is simplified to
\begin{eqnarray}
{\cal P}_\zeta \simeq \frac{1}{48\pi^2 g^2}\frac{V}{2\epsilon +\delta}.
\end{eqnarray}
In this case $\eta$ and $\dot J/HJ$ are related via
\begin{eqnarray}
\eta\simeq -\frac{2gU'}{VW}\frac{u'}{u}-\frac{\dot J}{HJ}.
\end{eqnarray}
Using this relation and Eq.~(\ref{dJHJ}),
one can eliminate $\eta$ in Eq.~(\ref{ns-extreme})
to express $n_s-1$ in terms of the $\phi$-dependent functions only.
The consistency relation is nothing but the standard one:
\begin{eqnarray}
r\simeq -8n_T.
\end{eqnarray}

On the other hand, if $J\simeq 6HXv$
then the power spectrum reduces to
\begin{eqnarray}
{\cal P}_\zeta \simeq \frac{\sqrt{6}}{128\pi^2 g^2}\frac{V}{2\epsilon +\delta}.
\end{eqnarray}
In this case $\eta$ and $\dot J/HJ$ are related via
\begin{eqnarray}
\eta\simeq-\frac{\epsilon}{2} -\frac{gU'}{VW}\frac{v'}{v}-\frac{1}{2}\frac{\dot J}{HJ},
\end{eqnarray}
which allows us to write $n_s-1$ in terms of the $\phi$-dependent functions only.
The consistency relation is given by the nonstandard one: 
\begin{eqnarray}
r\simeq -\frac{32\sqrt{6}}{9}n_T.
\end{eqnarray}

\subsection{Non-Gaussianity}

As with conventional potential-driven inflation models we expect small
non-Gaussianity in the models at hand. It is explicitly computed in the
Appendix.  In the limit $u\gg H\dot\phi v$, it turns out that the
equilateral $f_{\rm NL}$ is slow-roll suppressed. In the opposite limit,
$u\ll H\dot \phi v$, we find that the leading contribution is
independent of the slow-roll parameters:
\begin{eqnarray}
f_{\rm NL}\simeq \frac{235}{3888}.
\end{eqnarray}
However, in the special case $u/W\approx 2$
we have $c_s^2 \gg 1$. In this case $f_{\rm NL}$ can be as large as
\begin{eqnarray}
f_{\rm NL}\approx \frac{5}{81}c_s^2\gg 1.
\end{eqnarray}
This happens if $u\approx -6 H\dot\phi v$.

\section{Discussion}

We have presented a unified treatment of Higgs inflation models in the
context of the most general single-field inflation model with
second-order equations of motion, the generalized G-inflation, in which
all existing Higgs inflation models can be accommodated. This unified
approach also enabled us to find yet another class of Higgs inflation
models, running Einstein inflation.  Including this newly proposed model, we have studied five Higgs
inflation models on the same footing. Formulas for primordial fluctuations of the generalized Higgs
inflation were given, which would be quite useful to discuss and
discriminate the model from observations and experiments in the near
future such as the LHC and Planck satellite.

Although our analysis is applicable to a wide class of potential-driven
inflation models besides the SM Higgs inflation, as for the relevance to
the latter, it is important to analyze the stability of the theory at
the energy scale of inflation.  In fact, according to
\cite{EliasMiro:2011aa}, for the mass range of the SM Higgs particle
favored by the recent LHC result \cite{ATLAS,CMS} the parameter region
where the Higgs quartic coupling is positive and stable up to the
inflationary scale is disfavored\footnote{The possibility that the Higgs quartic 
coupling and its beta function vanish at the Planck scale has also been 
discussed in Ref.~\cite{planck0}.}, which might make all the Higgs
inflation models difficult or even impossible. However, as already
pointed out by the authors of Ref. \cite{EliasMiro:2011aa}, there are
still theoretical uncertainties on the beta functions and experimental
errors on the top and the Higgs masses, which make the conclusion
indecisive. In addition, the presence of higher order derivatives and
the coupling to the Einstein tensor may alter the results in the context
of generalized Higgs inflation. Therefore, proper analysis must be
performed in the framework of generalized G-inflation, which may
well improve the situation.\footnote{It is known that the introduction
of additional degrees of freedom relaxes the constraints
\cite{Lebedev:2012zw}.} We plan to study this issue by the time the
discovery of the Higgs particle is confirmed and its mass is fixed.
After completion of this study, we can answer the question of whether 
inflation can be explained within the SM or not\footnote{
Possible extensions of the SM accommodating Higgs inflationary scenarios
are discussed in Ref.~\cite{conf}. }.

\section*{Acknowledgments}

We would like to thank Matt Lake for a useful comment.  This work was
partially supported by JSPS Grant-in-Aid for Research Activity Start-up
No. 22840011 (T.K.); the Grant-in-Aid for Scientific Research
No. 23740195 (T.T.), No. 21740187 (M.Y.), and No. 23340058 (J.Y.); and the
Grant-in-Aid for Scientific Research on Innovative Areas No. 21111006
(J.Y.).

~\\
\newpage

\appendix

\section{Bispectrum}

In this appendix, we present the explicit formula for the bispectrum in
general potential-driven slow-roll inflation.  The bispectrum of the
curvature perturbation is defined as
\begin{eqnarray}
\langle \zeta_{\mathbf{k}_1} \zeta_{\mathbf{k}_2} \zeta_{\mathbf{k}_3}\rangle
=(2\pi)^3\delta(\mathbf{k}_1+\mathbf{k}_2+\mathbf{k}_3)B_\zeta(k_1,k_2,k_3).
\end{eqnarray}
Following the result of Refs. \cite{Gao:2011qe, DeFelice:2011uc, petel},
we find
\begin{widetext}
\begin{eqnarray}
B_\zeta&=&\frac{(2\pi)^4{\cal P}_\zeta^2}{4k_1^3k_2^3k_3^3}\left[
6{\cal C}_1\frac{(k_1k_2k_3)^2}{K^3}
+\frac{{\cal C}_2}{K}\left(2\sum_{i>j}k_i^2k_j^2-\frac{1}{K}\sum_{i\neq j}k_i^2k_j^3\right)
+{\cal C}_3\left(
\sum_ik_i^3+\frac{4}{K}\sum_{i>j}k_i^2k_j^2-\frac{2}{K^2}\sum_{i\neq j}k_i^2k_j^3
\right)
\right.
\nonumber\\
&&\qquad\left.
+\frac{{\cal C}_4}{K}\left(\sum_ik_i^4-2\sum_{i>j}k_i^2k_j^2\right)
\left(1+\frac{1}{K^2}\sum_{i>j}k_ik_j+\frac{3k_1k_2k_3}{K^3}\right)
\right], 
\end{eqnarray}
\end{widetext}
where $K := k_1+k_2+k_3$, each coefficient ${\cal C}_i$ is given by
% 
%\begin{widetext}
\begin{eqnarray}
{\cal C}_1&=&\frac{1}{c_s^2}-1+\frac{2(2-c_s^2)}{{\cal F}_S}
\frac{\dot\phi X}{H}v \nonumber \\ 
&\simeq& \frac{1}{c_s^2}-1+2(2-c_s^2)\frac{1-u/W}{4-u/W},
\\
{\cal C}_2&=&3\left(1-\frac{1}{c_s^2}\right),
\\
{\cal C}_3&=&
\frac{1}{2}\left(\frac{1}{c_s^2}-1\right),
\\
{\cal C}_4&=&-\frac{1}{c_s^2{\cal F}_S}\frac{\dot\phi X}{H}v
\simeq -\frac{1}{c_s^2}\frac{1-u/W}{4-u/W},
\end{eqnarray}
%\end{widetext}
and we have neglected the slow-roll suppressed contributions. Here, we
have also used the relations,
\begin{eqnarray}
&&{\cal F}_S=2g\epsilon+g\delta + \frac{\dot\phi X}{H}v,
\\
&&\frac{\dot\phi X}{H}v \simeq \frac{g}{3}\left(2\epsilon+\delta\right)
   \left(1-\frac{u}{W}\right).
\end{eqnarray}
It is found that
\begin{eqnarray}
{\cal C}_1, {\cal C}_2,{\cal C}_3,{\cal C}_4 \lesssim 1
\end{eqnarray}
for $c_s^2={\cal O}(1)$. However, one of the coefficients can be as
large as ${\cal C}_1\sim c_s^2\gg 1$ for $u/W\approx 2$.

Taking the equilateral configuration, $k_1=k_2=k_3$, the nonlinearity
parameter $f_{\rm NL}$ is given by
\begin{eqnarray}
f_{\rm NL}&=&\frac{5}{81}\left(
{\cal C}_1+6{\cal C}_2+\frac{51}{2}{\cal C}_3-\frac{13}{2}{\cal C}_4
\right) \nonumber \\ 
&\simeq&\frac{5}{243}\frac{\left( 1-u/W \right)^2\left( 99-43u/W \right)}
                           {\left( 4-u/W \right)^2\left( 2-u/W \right)}.
\end{eqnarray}

\end{document}